\documentclass[aps,prc,twocolumn,floatfix,showpacs,amsfonts,amssymb,amsmath]{revtex4}
\usepackage{graphicx}
\usepackage{rotating}

\begin{document}
\title{Comment on "New formulas for the ($-2$) moment of the photoabsorption cross section, $\sigma_{-2}$"}

\author{Peter von Neumann-Cosel}\email{vnc@ikp.tu-darmstadt.de}
\affiliation{Institut f\"ur Kernphysik, Technische Universit\"at
Darmstadt, 64289 Darmstadt, Germany}
\date{\today}

\begin{abstract}
Empirical formulas for the second inverse moment of the photoabsorption cross sections in nuclei are discussed in J.~N.~Orce, Phys.~Rev.~C~{\bf 91}, 064602 (2015).
In this Comment I point out that the experimental values used are systematically too small in heavy nuclei by about 5-10\% because of the neglection of the E1 strength below the neutron threshold.
Furthermore, combining recently deduced values of the polarizability in heavy and total photoabsorption data in light nuclei it is demonstrated that the mass number dependence of $\sigma_{-2}$ is sensitive to the volume and surface coefficients of the symmetry energy and parameters different to the ones chosen by Orce may be better suited.  
 \end{abstract}


\maketitle

In Ref.~\cite{orc15}, Orce discusses empirical relations of the second inverse moment of photabsorption cross sections ($\sigma_{-2}$)  in nuclei.
The results are based on the compilation of experimental $(\gamma,xn)$ cross sections by Dietrich and Berman \cite{die88}.
The quantity $\sigma_{-2}$ is proportional to the static electric dipole polarizability $\alpha_D$ and in practical units $\alpha_D$  (fm$^3$) $\simeq$ $\sigma_{-2}$ (mb/MeV). 
There is current interest into the polarizability of nuclei because it has been shown to be a measure of the neutron skin thickness in theoretical calculations based on energy density functionals (EDFs) \cite{rei10,pie12}.
Since there is a strong correlation in EDFs between the neutron skin thickness and the density dependence of the symmetry energy \cite{bro00}, important information on the equation of state of neutron matter can be derived (see Ref.~\cite{epja50} for a recent review).  

The two main results of Ref.~\cite{orc15} are a power law for the mass number ($A$) dependence 
\begin{equation}
\sigma_{-2} = 2.4 \times 10^{-3} A^{5/3} \, {\rm mb/MeV},
\label{eq:adep}
\end{equation}
which holds with a root mean square error of 30\% (6\% for $A \geq 60$) and the parameterization
\begin{equation}
\sigma_{-2} = \frac{1.8 \times 10^{-3} A^2}{A^{1/3} - 1.27} \, {\rm mb/MeV}.
\label{eq:sym}
\end{equation}
The latter relation takes into account the $A$ dependence of the symmetry energy term ($a_{\rm sym}$) in the Bethe-Weisz\"acker mass formula
\begin{equation}
a_{\rm sym}(A) = S_v \left ( 1 - \frac{\kappa}{A^{1/3}} \right ).
\label{eq:symth}
\end{equation}
leading to 
\begin{equation}
\sigma_{-2} = \frac{0.0518 \, A^2}{S_v (A^{1/3} - \kappa)} \, {\rm mb/MeV}.
\label{eq:sym2}
\end{equation}
Here $\kappa = S_s/S_v$, and $S_s$ and $S_v$ denote the surface and volume coefficients, respectively, and
the numerical coefficient is obtained from Migdal’s approach to the hydrodynamical model (cf.\ Eq. (1) in Ref. \cite{orc15}).
Equation (\ref{eq:sym}) was derived using values for $S_v$ and $\kappa$ from a recent fit of mass differences between isobaric nuclei \cite{tia14}.

In lighter nuclei, large deviations from relations (\ref{eq:adep},\ref{eq:sym}) were observed and explained to result from the restriction of the data to neutron decay channels which often represent a minor part of the photoabsorption cross sections only. 
It is the purpose of this Comment to (i) discuss the implications of low-energy contributions to $\sigma_{-2}$ neglected in the analysis of Ref.~\cite{orc15} and (ii) use total photoabsorption data (also available in light nuclei) for a test of Eq.~(\ref{eq:sym}).  

The electric dipole response below the particle thresholds is a subject of current interest.
In medium-mass and heavy nuclei with neutron excess a resonance-like structure is observed often termed Pgmy Dipole Resonance (PDR) and suggested to result from an oscillation of the excess neutrons forming a skin against a $N \approx Z$ core \cite{paa07}.
Typically, the PDR contribution to the summed photoabsorption cross section is small, although in exotic neutron-rich nuclei values up to about 5\% have been reported \cite{adr05,wie09}. 
However, the systematics of the PDR strength are not well understood \cite{sav13}, mainly because the results from the most widely used experimental method, nuclear resonance fluorescence, show large uncertainties due to unknown branching ratios to excited states.

Recently, a new method to measure the complete E1 response from about 5 to 25 MeV based on relativistic Coulomb excitation in $(p,p')$ scattering at very forward angles has been developed.
Application to $^{208}$Pb \cite{pol12} and $^{120}$Sn \cite{kru15} shows cumulated E1 strengths exhausting about 1.5\% and 2.5\%, repectively, of the energy-weighted sum rule below neutron threshold.
Because of the inverse energy weighting, the contributions to the polarizability amount to about 6.5\% for $^{208}$Pb and about 9\% for $^{120}$Sn.
It can be expected that this low-energy contribution appears in all medium-heavy and heavy nuclei with neutron excess and depends on the threshold energy (e.g., $B_n = 7.33$ MeV for $^{208}$Pb and 9.10 MeV for $^{120}$Sn).
Comparable amounts of E1 strength below threshold have been found in $^{90}$Zr \cite{iwa12} and $^{138}$Ba \cite{ton10}.
In deformed nuclei, the corrections due to E1 strength below the threshold may even be larger because of the broadening of the GDR \cite{kru14}.
Overall one may expect a systematic upward correction of the $\sigma_{-2}$ values used in Ref.~\cite{orc15} of the order of 5-10\%. 

\begin{figure}[t]
\includegraphics[width=8.5cm]{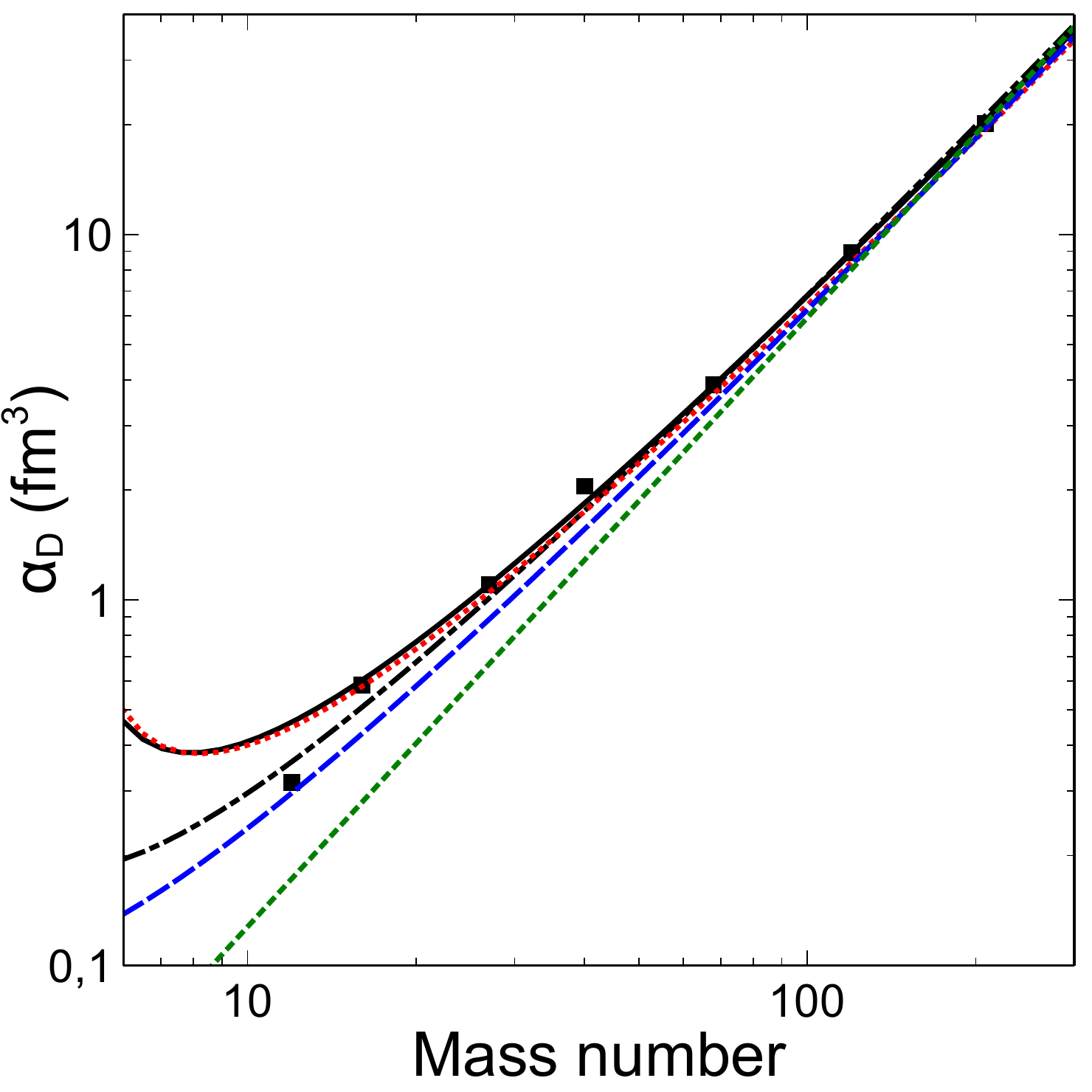} 
\caption{\label{figure}
(Color online). Full Squares: $\sigma_{-2}$ data deduced from relativistic Coulomb excitation experiments for $^{208}$Pb \cite{tam11}, $^{120}$Sn \cite{has15}, and $^{68}$Ni \cite{ros13,roc15}  and from total photoabsorption experiments  for $^{40}$Ca, $^{27}$Al, $^{16}$O, and $^{12}$C \cite{ahr75}.
Error bars are partly smaller than the symbol size.
Short-dashed (green) and long-dashed (blue) lines: Empirical formulas (\ref{eq:adep}) and (\ref{eq:sym}). 
Dotted (red) line: Equation (\ref{eq:sym2}) using  symmetry energy parameters of Ref.~\cite{ste05}.
Dashed-dotted and full black lines: Fit to Eq.~(\ref{eq:sym2}) including and excluding the $^{12}$C data point.} 
\end{figure}
The combination of the data from relativistic Coulomb excitation with $(\gamma,xn)$ cross sections allows a precise extraction of the dipole polarizability of $^{120}$Sn \cite{has15} and $^{208}$Pb \cite{tam11}. 
Additionally, {\it total} photoabsorption data in light nuclei over a wide energy range are reported in Ref.~\cite{ahr75}.
Although natural targets were used in these measurements, a single isotope is most abundant for each element.
Thus, the results are representative for $^{12}$C, $^{16}$O, $^{27}$Al, and $^{40}$Ca.
(Note that Ref.~\cite{ahr75} also provides data for $^7$Li and $^9$Be but the hydrodynamical picture is highly questionable and corrections due to the magnetic polarizability are large \cite{knu81}) for these very light nuclei).  
Magnetic contributions to the $\sigma_{-2}$ values have been separated for $^{120}$Sn \cite{kru15} and $^{208}$Pb \cite{pol12} and can generally be neglected for $A \geq 12$.

This set of data including a recent result for $^{68}$Ni \cite{ros13} with corrections for unobserved strength \cite{roc15} is shown in Fig.~\ref{figure} as full squares.
The value for $^{40}$Ca ($\sigma_{-2} = 2.05(10)$ mb/MeV) differs from Table II in Ref.~\cite{ahr75} because the data  with very coarse energy binning in the GDR energy region was replaced by subsequent results with finer energy steps by the same group \cite{ahr85}, cf.~Ref.~\cite{exfor}.  
The data cover a wide range of mass numbers and thus permit a test of Eqs.~(\ref{eq:adep}) and (\ref{eq:sym}) shown as short-dashed (green) and long-dashed (blue) lines in Fig.~\ref{figure}, respectively.

The experimental results are systematically larger than Eq.~(\ref{eq:adep}) as expected from the above arguments. 
The deviation increases towards smaller mass numbers.
Equation (\ref{eq:sym})  leads to similar results for heavy nuclei.
The description for lighter masses is improved but still unederstimates the data except for $^{12}$C.
The numerical coefficents in Eq.~(\ref{eq:sym}) stem from the mass dependence of the symmetry energy  [Eq.~(\ref{eq:symth})] using the parameters of Ref.~\cite{tia14} ($S_v = 28.3$ MeV, $\kappa = 1.27$).
Similar values have been reported by Ref.~\cite{die05}.
However, alternative parameters have been derived e.g.\ in Refs.~\cite{dan03,ste05}.
While the value of $S_v$ is fairly consistent in all models, larger values of $\kappa$ are obtained in the latter approaches.
The dotted (red) line in Fig.~\ref{figure} uses parameters of Ref.~\cite{ste05} ($S_v = 27.3$ MeV, $\kappa = 1.68$) and provides a good description of the data both in absolute magnitude as well as reproducing the $A$ dependence with the exception of $^{12}$C.  
An alternative parameter set ($S_v = 24.1$ MeV, $\kappa = 0.545$) discussed in Ref.~\cite{ste05} completely fails to describe the data. 

One can also perform a free fit to Eq.~(\ref{eq:sym2}).
The result depends crucially on the inclusion (black solid line) or exclusion (black dashed-dotted line) of the $^{12}$C data point.  
In the former case, the results [$S_v = 23.5(7)$ MeV, $\kappa = 1.41(5)$, $\chi^2/{\rm dof} = 5.7$] are closer to Eq.~(\ref{eq:sym}).
The latter analysis without the $^{12}$C result provides a better fit to the data ($\chi^2/{\rm dof} = 1.3$) with parameters [$S_v = 25.6(8)$ MeV, $\kappa = 1.66(5)$] similar to those of Ref.~\cite{ste05}. 
These examples illustrate the importance of studying the experimental systematics of $\sigma_{-2}$  (i.e., the polarizability) over a wide mass range.
Despite the limitations of the underlying approach neglecting structure effects one can expect relevant information on the volume and surface coefficients and thus the density dependence of the symmetry energy.  

\begin{acknowledgments}
This work was supported by the DFG under contract SFB 1245.
\end{acknowledgments}


\begin{thebibliography}{abc99x}

\bibitem{orc15}
J. N. Orce,
%
Phys. Rev. C {\bf 91}, 064602 (2015).

\bibitem{die88}
S. S. Dietrich and B. L. Berman, Atom. Data Nucl. Data Tables {\bf 38}, 199 (1988).

\bibitem{rei10}
%
P.-G. Reinhard and W. Nazarewicz, Phys. Rev. C {\bf 81}, 051303(R) (2010).

\bibitem{pie12} 
%
J. Piekarewicz {\it et al.,}, Phys. Rev. C {\bf 85}, 041302(R) (2012).

\bibitem{bro00} 
%
B. A. Brown, Phys. Rev. Lett. {\bf 85}, 5296 (2000).

\bibitem{epja50}
%
{\it Topical Issue on Nuclear Symmetry Energy}, edited by Bao-An Li, A. Ramos, G. Verde and I. Vida\~{n}a, Eur. Phys. J. A {\bf 50(2)} (2014).

\bibitem{tia14}
J. Tian, H. Cui, K. Zheng, and N. Wang, Phys. Rev. C {\bf 90}, 024313 (2014).

\bibitem{paa07}
N. Paar, D. Vretenar, E. Khan, and G. Col\`{o},
%
Rep. Prog. Phys. {\bf 70}, 691 (2007).

\bibitem{adr05}
P. Adrich {\it et al.},
%
Phys. Rev. Lett. {\bf 95}, 132501 (2005).

\bibitem{wie09}
O. Wieland {\it et al.},
%
Phys. Rev. Lett. {\bf 102}, 092502 (2009).

\bibitem{sav13}
D. Savran, T. Aumann, and A. Zilges,
%
Prog. Part. Nucl. Phys. {\bf 70}, 210 (2013).

\bibitem{pol12}
%
I. Poltoratska {\it et al.}, 
%
Phys. Rev. C {\bf 85}, 041304(R) (2012).

\bibitem{kru15}
%
A. M. Krumbholz {\it et al.}, 
%
Phys. Lett. B {\bf 744}, 7 (2015).

\bibitem{iwa12}
C. Iwamoto {\it et al.},
%
Phys. Rev. Lett. {\bf 108}, 262501 (2012).

\bibitem{ton10}
A. P. Tonchev, S. L. Hammond, J. H. Kelley, E. Kwan, H. Lenske, G. Rusev, W. Tornow, and N. Tsoneva,
%
Phys. Rev. Lett. {\bf 104}, 072501 (2010).

\bibitem{kru14}
A. Krugmann, D. Martin, P. von Neumann-Cosel, N. Pietralla, and A. Tamii,
%
EPJ Web of Conferences {\bf 66}, 02060 (2014).

\bibitem{has15}
T. Hashimoto {\it et al.}, 
%
Phys. Rev. C {\bf 92}, 031305(R) (2015).

\bibitem{tam11}
A. Tamii {\it et al.,}
%
 Phys. Rev. Lett. {\bf 107}, 062502 (2011).

\bibitem{ahr75}
J. Ahrens {\it et al.},
%
 Nucl. Phys. {\bf A251}, 479 (1975).
 
 \bibitem{knu81}
 W. Kn\"{u}pfer and A. Richter,
 %
 Phys. Lett. B {\bf 107}, 325 (1981).

\bibitem{ros13}
D. M. Rossi {\it et al.},
%
Phys. Rev. Lett.{\bf 111}, 242503 (2013).

\bibitem{roc15}
X. Roca-Maza, X. Vi\~nas, M. Centelles B. K. Agrawal, G. Col\`o, N. Paar, J. Piekarewicz, and D. Vretenar,
%
Phys. Rev. C {\bf 92}, 064304 (2015).

\bibitem{ahr85}
J. Ahrens,
%
Nucl. Phys. {\bf A446}, 229c (1985).

\bibitem{exfor}
N. Otuka {\it et al.},
%
Nucl. Data Sheets {\bf 120}, 272 (2014);
https://www-nds.iaea.org/exfor/exfor.htm.

\bibitem{die05}
A. E. L. Dieperink and D. van Neck, 
%
J. Phys.: Conf. Ser. {\bf 20}, 160 (2005).

\bibitem{dan03}
P. Danielewicz,
%
Nucl. Phys. {\bf A727}, 233 (2003).

\bibitem{ste05}
A. W. Steiner, M. Prakash, J. M. Lattimer,  and P. J. Ellis,
%
Phys. Rep. {\bf 411}, 325 (2005) .

\end{thebibliography}
\end{document}